\documentclass[aps,pra,twocolumn,groupedaddress]{revtex4}

\newif\ifpdf
\ifx\pdfoutput\undefined
\pdffalse 
\else
\pdfoutput=1 
\pdftrue
\fi

\ifpdf
\usepackage[pdftex]{graphicx}
\else
\usepackage{graphicx}
\fi
\usepackage{hyperref}

\begin{document}

\ifpdf
\DeclareGraphicsExtensions{.pdf, .jpg}
\else
\DeclareGraphicsExtensions{.eps,.ps, .jpg}
\fi

\def\hslash{\hbar}
\def\imag{i}
\def\grad{\vec{\nabla}}
\def\div{\vec{\nabla}\cdot}
\def\curl{\vec{\nabla}\times}
\def\DDt{\frac{d}{dt}}
\def\ddt{\frac{\partial}{\partial t}}
\def\ddx{\frac{\partial}{\partial x}}
\def\ddy{\frac{\partial}{\partial y}}
\def\lap{\nabla^{2}}
\def\divv{\vec{\nabla}\cdot\vec{v}}
\def\gradS{\vec{\nabla}S}
\def\vvec{\vec{v}}
\def\wc{\omega_{c}}
\def\<{\langle}
\def\>{\rangle}
\def\Tr{{\rm Tr}}
\def\Csch{{\rm csch}}
\def\Coth{{\rm coth}}
\def\Tanh{{\rm tanh}}
\def\g2{g^{(2)}}


\title{Exciton Dissociation Dynamics in Model Donor-Acceptor 
Polymer Heterojunctions: I. Energetics and Spectra}
\author{Eric R. Bittner}
\email[email:]{bittner@uh.edu}

\affiliation{Department of Chemistry and Center for Materials Chemistry, 
University of Houston \\ Houston, TX 77204}

\author{John Glenn Santos Ramon}

\affiliation{Department of Chemistry and Center for Materials Chemistry, 
University of Houston \\ Houston, TX 77204}

\author{Stoyan Karabunarliev}

\affiliation{Department of Chemistry and Center for Materials Chemistry, 
University of Houston \\ Houston, TX 77204}

\date{\today}

\begin{abstract}

 In this paper we consider the essential electronic excited states in parallel chains
 of semiconducting polymers that are currently being explored for photovoltaic and 
 light-emitting diode applications.   In particular, we focus upon various type II
donor-acceptor heterojunctions and explore the relation between the exciton binding
energy to the band off-set in determining the device characteristic of a particular 
type II heterojunction material.   As a general rule, when the exciton binding energy is 
greater than the band off-set at the heterojunction, the exciton will remain the lowest energy
excited state and the junction will make an efficient light-emitting diode.  On the other hand, 
if the off-set is greater than the exciton  binding energy, either the electron or hole
can be transferred from one chain to the other.   Here we use a two-band exciton
to predict the vibronic absorption and emission 
spectra of model polymer heterojunctions.   Our results underscore the role of
vibrational relaxation and suggest that intersystem crossings may play some part in the  
formation of charge-transfer states following photoexcitation in certain cases. 

\end{abstract}

\maketitle

\section{Introduction\label{intro}}

\begin{figure}[t]
\includegraphics[width=\columnwidth]{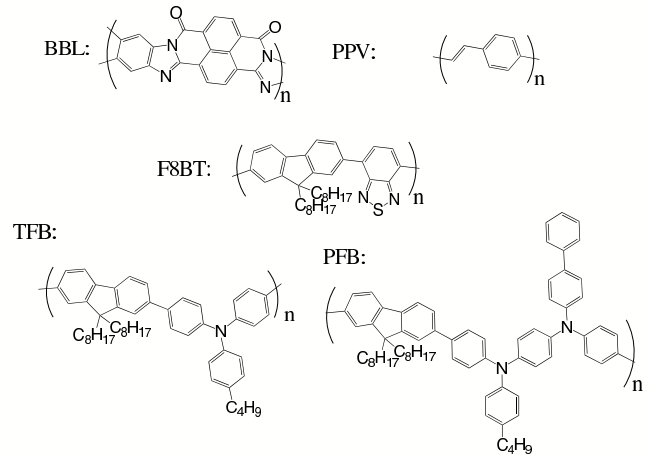}
\caption{Chemical structure of the various conjugated polymers considered in this paper. }\label{chains}
\end{figure}

Organic semiconducting polymers are currently of broad interest 
as potential low-cost materials for photovoltaic and light-emitting display
applications.   While research into the fabrication, chemistry, and 
fundamental physics  of these materials continues to advance,  devices 
based upon organic semiconductors have begun to appear in the market place. 
Composite materials fabricated by mixing semiconducting polymers with 
different electron and hole accepting properties have been used to make highly efficient 
solar cells and photovoltaic cells.\cite{tang:183,granstrom:257,bach:583} The advantage gained by mixing materials is that
when a conjugated polymer material absorbs light, the primary species produced is
an exciton, i.e. an electron/hole pair bound by Coulombic attraction, rather than 
free charge carriers.    As a result, a single layer conventional Schottky-barrier 
cell of a molecular semiconductor sandwiched between two metal electrodes
has poor power conversion and charge collection efficiencies.   In 
composite materials, the driving force for charge separation is due to the mismatch
between the HOMO and LUMO levels of the blended materials.  This creates
a band off-set at the interface between the two materials.   A type I heterojunction
is when the HOMO and LUMO levels of one of the materials both lie within the 
HOMO-LUMO gap of the other.   A type II heterojunction occurs when both the 
HOMO and LUMO of one material are simultaneously shifted up or down relative to the
HOMO and LUMO levels of the other material.   By and large, organic semiconductor 
heterojunctions fall into the type II category.\cite{shockley:1950}  The chemical structure of some of the more 
important materials for device applications are shown in Fig.~\ref{chains}.

Photocurrent production within a type II material hinges upon its ability to dissociate
the bound electron/hole pairs produced by photoexcitation.  To a first approximation, 
the dissociation threshold is determined by the ratio of the exciton binding energy, $\varepsilon_B$, to the 
band off-set at the heterojunction, $\Delta E$.  If $\Delta E > \varepsilon_B$, the exciton 
will be energetically unstable and can undergo fission to form charge-separated states and 
eventually free charge carriers. 
On the other hand,  when $\Delta E < \varepsilon_B$, the exciton remains the favorable 
species.   For typical organic semiconductors, $\varepsilon_B\approx 0.5 - 0.$6 eV. 
In heterojunctions composed of 
poly(benzimidazobenzophenanthroline ladder) (BBL) and 
poly(1,4-phenylenevinylene) (PPV),  $\Delta E$ is considerably greater than $ \varepsilon_B$\cite{alam:4647}.
As a result, recent reports of devices fabricated from nanoscale layers of BBL and PPV or its derivative 
poly(2-methoxy-5(2''-ethyl-hexyloxy)-1,4-phenylenevinylene) (MEH-PPV) indicate 
very efficient photoinduced charge transfer and large photovoltaic power conversion efficiencies under 
white light illumination.\cite{yu:5243,halls:498,halls:5721}

Surprisingly, type II materials can also be used to produce highly efficient light-emitting diodes (LEDs).
In demixed blends of poly(9,9-dioctylfluorene-co-benzothiadiazole) (F8BT) and 
poly(9,9-dioctylfluorene-co-N-(4-butylphenyl) diphenylamine) (TFB), the exciton is the stable species and 
fabricated devices show excellent LED performance\cite{morteani:1708}.    
While chemically similar to TFB, blends of F8BT with 
poly(9,9-dioctylfluorene-co-bis-N,N-(4 -butylphenyl)-bis-N,N-phenyl-1,4-phenylenediamine) 
(PFB) exhibit very poor LED performance.    Changing the diphenylamine to a phenylenediamine 
within the co-polymer shifts the position of the HOMO energy level by +0.79 eV relative to 
the HOMO level of TFB. (Table~\ref{table1}) and destabilizes the exciton. 

 In this paper 
  we investigate the role of interchain coupling  and lattice reorganization in 
determining the essential states model polymer semiconductor systems consisting of parallel 
chains of PPV:BBL,  TFB:F8BT,  and  PFB:F8BT.   
In the case of the PPV:BBL heterojunction, the exciton binding energy is smaller than the 
band off-set leading to electron transter from the photoexcited PPV to the BBL chain. 
In the case of TFB:F8BT and PFB:F8BT junctions, the band off-sets are close to the 
exciton binding energy and we see considerable mixing between intramolecular 
excitonic states with intermolecular charge-transfer or exciplex states.  
Moreover, in TFB:F8BT, avoided crossing between Born Oppenhemier potential energy surfaces
may open channels for exciton regeneration that are thermally accessible 
from the exciplex states as recently reported by Morteani {\em et al.}\cite{morteani:247402}.
In a subsequent paper we  will report upon our 
simulations of  the charge-injection and photoexcitation  dynamics within these systems. \cite{bittner05b}

 \begin{table}[b]
  \caption{Band centers and reported HOMO and LUMO levels for various polymer species.  Parenthesis indicate the 
modulation of the intramolecular valence and conduction band site energies. \label{table1}}
 \begin{ruledtabular}
 \begin{tabular}{c|cccc}
Molecule    &   $\varepsilon_e$ & $\varepsilon_h $  & HOMO & LUMO  \\
\hline
PPV           &      2.75 eV            &  -2.75 eV              &  -5.1 eV$^1$   & -2.7 eV$^1$  \\
BBL            &      1.45         &  -3.55          &  -5.9$^1$ & -4.0$^1$  \\
F8BT            &     1.92 (2.42,1.42)    &  -3.54 (-3.04,-4.04)         & -5.33$^2$  & -3.53$^2$\\
PFB           &        3.16  (3.36,2.96) &  -2.75    (-2.55,-2.95))& -5.1$^2$   &-2.29$^2$\\
TFB            &        3.15  (3.35,2.95) &  -2.98     (-2.78,-3.18) &  -5.89$^2$   & -2.30$^2$ \\
 \end{tabular}

 \end{ruledtabular}
 
 1.) M. M. Alam and S. A. Jenekhe, Chem. Mater. {\bf 16}, 4647 (2004). 2.) A. C. Morteani, P. Sreearunothai, L. M. Herz, R. H. Friend, and C. Silva, 
 Phys. Rev. Lett. {\bf 92},  247402 (2004).
 \end{table}

\section{Theoretical Approach}
Our basic description is derived starting from a model for the on-chain electronic excitations
of a single conjugated polymer chain.\cite{karabunarliev:4291,karabunarliev:10219,karabunarliev:057402}  This model accounts for the coupling of excitations within the $\pi$-orbitals of
a conjugated polymer to the lattice phonons using localized valence and 
conduction band Wannier functions ($|\overline{h}\>$ and $|p\>$) 
to describe the
$\pi$ orbitals and two optical phonon branches to 
describe the bond stretches and torsions of the
the polymer skeleton. 

\begin{eqnarray}
H &=& \sum_{{\bf m n}} (F^\circ_{\bf mn} +V_{\bf mn} )A_{\bf m}^\dagger A_{\bf n} 
\nonumber \\
&+& \sum_{{\bf nm}  i\mu }\left(\frac{\partial F^\circ_{\bf nm}}{\partial q_{i\mu}} \right)
A_{\bf n}^\dagger A_{\bf m}  q_{i\mu} \nonumber \\
&+& \sum_{i\mu}\omega_\mu^2(q_{i\mu}^2 + \lambda_\mu q_{i\mu} q_{i+1,\mu}) + p_{i\mu}^2
\end{eqnarray}
where  $A^\dagger_{\bf n}$ and $A_{\bf n}$ are Fermion operators that act 
upon the ground electronic state $|0\rangle$ to
create and destroy electron/hole configurations
$|n\>  = | \overline{h}p \>$ with positive hole in the valence band Wannier function localized at 
$h$ and an electron in the conduction band Wannier function $p$.  Finally,  $q_{i\mu}$ and $p_{i\mu}$
correspond to lattice distortions and momentum components in the $i$-th site and $\mu$-th optical phonon branch.
%
Except as noted below, our model and approach 
is identical to what we have 
described in our earlier publications.
\cite{karabunarliev:4291,karabunarliev:10219,karabunarliev:3988}

Since we are dealing with multiple polymer chains, we consider both interchain and 
intrachain single particle contributions.  For the intrachain terms, we use the 
hopping terms and site energies derived for isolated polymer chains of a given 
species, $t_{i,||}$, where our notation denotes the parallel hopping term for the $i$th
chain ($i = 1,2$).   For PPV and similar conjugated polymer species, these are 
approximately $0.5$eV for both valence and conduction $\pi$ bands. 




We include two intramolecular optical phonon branches which correspond roughly to 
the high-frequency C=C bond stretching modes within a given repeat unit and a second low-frequency
mode, which in the case of PPV are taken to represent the phenylene torsional modes.  
The electron-phonon couplings are assumed to be transferable between the various chemical species. 
Since the modes are assumed to be intramolecular, we do not include interchain couplings
in the phonon Hessian matrix.

Upon transforming $H$ into the diabatic representation by diagonalizing the electronic terms at $q_{i\mu} = 0$, 
we obtain a series of vertical excited states $|a_\circ\>$ 
with energies, $\varepsilon_a^\circ$ and normal modes, $Q_\xi$ with frequencies, 
$\omega_\xi$.  (We will assume that the sum over $\xi$ spans all phonon branches). 
\begin{eqnarray}
H &=& \sum_a \varepsilon_a^\circ |a_\circ\> \<a_\circ|+ \sum_{ab\xi}g^\circ_{ab\xi}q_\xi(|a_\circ\>\<b_\circ| +|b_\circ\>\<a_\circ|  ) \nonumber \\
&+& \frac{1}{2}\sum_\xi (\omega_\xi^2 Q_\xi^2 + P_\xi^2 ).
\end{eqnarray}
The adiabatic or relaxed states can be determined then by iteratively minimizing $\varepsilon_a(Q_\xi)  = \<a|H|a\>$ according to the self-consistent equations
\begin{eqnarray}
\frac{d\varepsilon_a(Q_\xi) }{dQ_\xi}  = g_{aa\xi} + \omega_\xi^2Q_\xi  = 0.
\end{eqnarray}

Thus, each diabatic potential surface for the nuclear lattice motion is given by
\begin{eqnarray}
\varepsilon_a(Q_\xi) = \varepsilon_a + \frac{1}{2}\sum_\xi \omega_\xi^2 (Q_\xi - Q_{\xi}^{(a)})^2.
\end{eqnarray}
While our model accounts for the distortions in the lattice due to electron/phonon coupling, we do not account for any adiabatic change in the phonon force constants 
within the excited states.   

While following along a relaxation pathway,  crossing between diabatic states can occur. 
In order to insure maximum correlation between the relaxed state, $\psi_b(Q_\xi)$
and its parent vertical state $\psi_{a\circ}(0)$, we compute the inner product, 
$S_{ab} = \<\psi_{a\circ}(0)|\psi_b(Q_\xi)\>$ and select the state with the greatest overlap 
with the parent diabatic state.   This allows us to follow a given state and correlate 
the adiabatic states with the parent vertical states.  Figures~\ref{ppv-bbl-correlation}
and \ref{tfb-pfb-f8bt-correlation} are examples of this. The dashed lines connect the 
parent vertical state with the final relaxed adiabatic state.  We will comment upon these 
figures later in this paper.

\begin{figure}[h]
\includegraphics[width=\columnwidth]{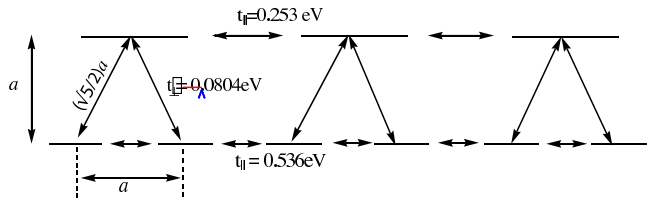}
\caption{Modified ladder coupling scheme for BBL:PPV interchain interactions.  
 The parallel and perpendicular hopping integrals ($t_{||}$ and $t_\perp$) for the two 
 chains and the separation distances (in units of lattice spacing $a$) are 
 indicated on this figure. For the polymers
under consideration,  the top chain represents the BBL, chain while the
sites correspond to the PPV chain.}\label{ladder}
\end{figure}

Since we are dealing with inter-chain couplings we make the following 
set of assumptions.  First, the single-particle coupling between chains is 
expected to be small compared to the intramolecular coupling.   For this, we assume that
the perpendicular hopping integral $t_\perp = 0.01 eV$.  This is consistent with 
LDF calculations performed by Vogl and Campbell and with the 
$t_\perp \approx 0.15 f_1$ estimate used in an 
earlier study of interchain excitons by Yu {\em et al}.\cite{vogl:12797,yu:8847}
Furthermore, we assume that the $J(r)$, $K(r)$, and $D(r)$ two-particle interactions 
depend only upon the linear distance between two sites, as in the intrachain case. 
Since these are expected to be weak given that the interchain separation, $d$, is
taken to be some what greater than the inter-monomer separation.   The third approximation 
that we make takes into account the difference in size of the repeat units of the
polymer chains.   As indicated in Fig.~\ref{chains}, the monomeric unit in BBL is 
roughly two times the size as a PPV monomeric unit.  Thus, for the PPV:BBL 
junction we adopt the modified ladder scheme show in Fig.~\ref{ladder}.
 
Finally, the most important assumption that we make is that the site energies for the electrons and 
holes for the various chemical species can be determined by comparing the relative HOMO and LUMO
energies to PPV.  These are listed in Table ~\ref{table1}.
For example, the HOMO energy for PPV (as determined by its ionization potential) is 
-5.1eV.   For BBL, this energy is -5.9 eV.  Thus, we assume that the 
$\overline{f}_o$ for a hole on a BBL chain is 0.8 eV lower than $\overline{f}_o$ for PPV at -3.55eV. 
 Likewise for the conduction band.   The LUMO energy of PPV is -2.7eV and that of BBL is -4.0eV. 
 Thus, we shift the band center of the BBL chain  1.3eV lower than then PPV conduction band center 
 to 1.45eV.    For the F8BT, TFB, and PFB chains, we adopt a similar scheme as discussed below.  
 The site energies and transfer integrals used throughout are indicated in Table~\ref{table1}.
 We believe our model  produces a reasonable estimate of the band off-sets in the PN-junctions 
 formed at the interface between these semiconducting polymers.

\section{Excitonic vs. Charge Separated States in Donor/Acceptor Polymer Junctions}
\subsection{Electronic States of Isolated Chains}

Before moving on to consider the electronic states of the heterojunction systems, 
we make a brief examination of the CI states of the isolated chains.  
The relevant data for the vertical and adiabatic excitonic and charge-separated states are
given in Table~\ref{table3}.    Since we have assumed a symmetric band structure, $f_1 = -\overline{f}_1$,  the electron/hole wavefunctions will have even or odd symmetry under spatial 
inversion.  Hence the total hamiltonian can be block diagonalized into even and odd parity blocks.
The even parity states we term XT states since these are optically coupled to the ground state
and are dominated by geminate electron/hole configurations.  On the other hand, the odd-parity
states are purely charge-separated states  (CT) which 
 possess no geminate electron/hole configurations, are optically coupled weakly at best to the 
ground state.   We define the exciton binding energy to be the energy difference between the 
lowest energy XT and CT states.  In each of the model systems considered, we obtain an
exciton binding energy between 0.5 and 0.65V.

\begin{table}[b]
  \caption{Vertical and Adiabatic singlet exciton energies for isolated polymer chains.  \label{table3}}
 \begin{ruledtabular}
 \begin{tabular}{c|cccccc}
Molecule    &  length &$\varepsilon_{EX}^\circ$ &  $\varepsilon_{EX}$  &  $\varepsilon_{CT}^\circ$ &  $\varepsilon_{CT}$ &$ \varepsilon_{B}$  \\
\hline
PPV           & 10  &2.56 eV& 2.37 eV & 3.01 eV & 2.877 eV  &  0.51 eV  \\
BBL            &  5 & 3.34   &3.07   &3.80      & 3.64         & 0.58\\
F8BT          & 10    &    2.28  &  2.12 & 2.79        &  2.71   & 0.59  \\
PFB            &    10 &  2.94  & 2.744  & 3.37   & 3.23  &0.49\\
TFB            &    10 &  3.15 &  2.96&3.59     &  3.45  & 0.49 \\
 \end{tabular}

 \end{ruledtabular}
   $\varepsilon_{XT}^\circ$: vertical symmetric exciton energy,\\
 $\varepsilon_{XT}$:   adiabatic symmetric exciton energy, \\
  $\varepsilon_{CT}^\circ$: vertical   asymmetric exciton energy, \\
  $\varepsilon_{CT}$: adiabatic asymmetric exciton energy,\\
$ \varepsilon_{B}$: adiabatic exciton binding energy.  
  \end{table} 

\begin{figure}[t]
\includegraphics[width=\columnwidth]{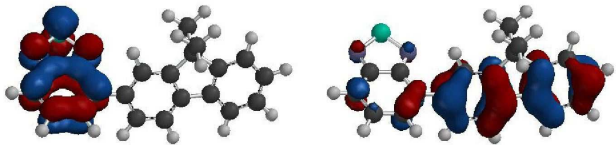}
\caption{Semiempirical (PM3) LUMO (left) and HOMO (right) orbitals for FBT monomer.}\label{fbt-orbitals}
\end{figure}

A  uniform site model for F8BT, TFB, and PFB, may be 
a gross simplification of the physical systems.  For example, recent 
semi-empirical CI calculations by Jespersen, et al\cite{jespersen:12613} 
indicate that the lowest energy singlet excited state of F8BT 
consists of alternating positive and negative regions corresponding to the electron 
localized on the benzothiadiazole units and the hole localized on the fluorene units.  These
are consistent with a previous study by Cornil {\em et al.}\cite{cornil:6615} which places the LUMO
on the benzothiadiazole units. Cornil {\em et al.}\cite{cornil:6615} 
also report the HOMO and LUMO levels of the 
isolated fluorene and benzothiadiazole
 Ref.~\cite{cornil:6615} indicating a $\Delta = 1.56$ eV
off-set  between the fluorene and benzothiadiazole  LUMO levels and a 0.66 eV off-set between
 the fluorene and benzothiadiazole HOMO 
energy levels.   Similarly, PM3 level calculations at the optimized geometry indicate a 
LUMO offset of 1.48eV and a HOMO offset of 0.8eV.  The HOMO and LUMO orbitals for
FBT (where we replaced the octyl side chains in F8BT with methyl groups) are shown in  Fig.~\ref{fbt-orbitals}. 
This clearly indicates the localization of the HOMO and LUMO wave functions on the co-polymer 
units. 

We can include this alternation into our model by modulating the site energies of the 
F8BT chain\cite{karabunarliev:10219}.   Thus, in F8BT we include a 0.5 eV modulation
of both the valence and conduction band site energies relative to the band center.
 Table~\ref{table1}.    Hence, the fluorene site energies are at 2.42eV and  -3.04eV for the conduction and 
valence band while the benzothiadiazole site energies are 1.42eV and -4.04eV. 
This results in a  shift in the excitation energy to 0.28 eV relative to the unmodulated
 polymer and a 0.09 eV increase in the exciton binding energy.  
Furthermore, the absorption spectrum consists of two distinct peaks at 2.14eV  (563nm) 
and 4.4eV (281nm) which are more or less on par with the  2.77eV (448nm) $S_o\rightarrow S_1$
and the 4.16eV (298 nm)  $S_o\rightarrow S_9$ transitions computed by Jespersen 
{\em et al.}~\cite{jespersen:12613}
and observed  at 2.66eV and 3.63eV by Stevens, {\em et al.} ~\cite{stevens:165213}. 

\begin{figure}[b]
\includegraphics[width=\columnwidth]{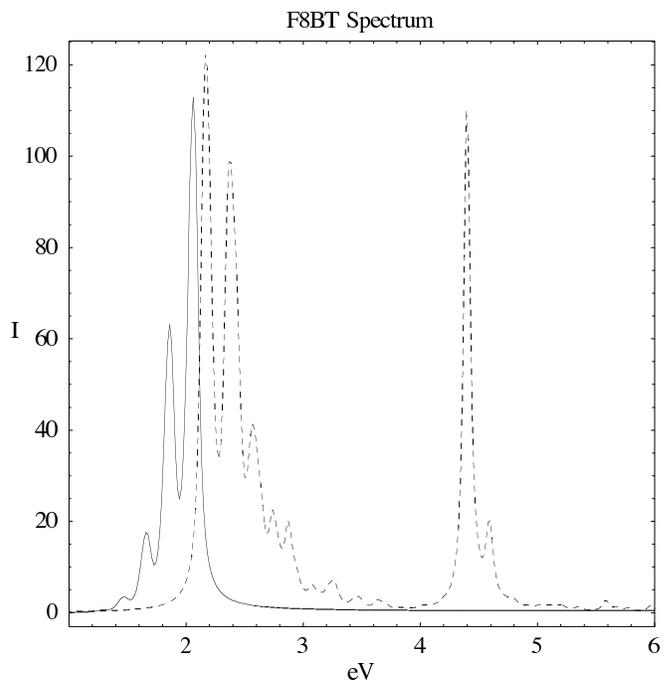}
\caption{ Absorption (dashed) and vibronic 
emission (solid) spectra for the modulated F8BT chain. The vibronic fine-structure
in each peak is due to the C=C bond-stretching phonon branch.} \label{f8bt-mod-spectrum}
\end{figure}

\begin{figure}[t]
\includegraphics[width=0.7\columnwidth]{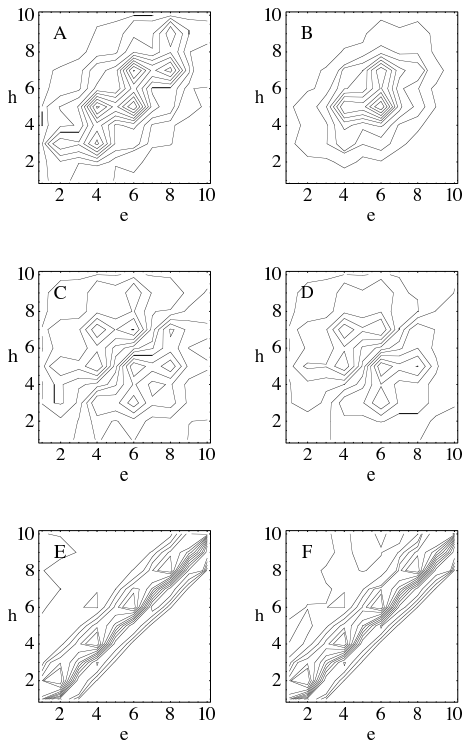}
\caption{
Vertical (A,C,D) and relaxed (B,D,F) electron/hole densities for 
modulated F8BT chain (F8BT-mod). A,B) Exciton, C,D.)Asymmetric Charge-transfer state,  
E,F) $\pi^*$ Exciton.  States A,B and E, F correspond to the $S_1$ CT  and $S_9$ $\pi\rightarrow \pi^*$ 
states reported in Ref.~\cite{jespersen:12613}.} \label{f8bt-mod-states}
\end{figure}

The charge-separated character of the lowest singlet excited state of the F8BT-mod chain is 
readily apparent in Fig.~\ref{f8bt-mod-states} A and B.  
This is the state which give rise to the absorption maxima at 2.14 eV and is the only 
state present in the emission spectra.    The odd-numbered sites are fluorene-sites and the 
even-numbered sites are benzothiadiazole sites. 
 The alternating maxima in Fig.~\ref{f8bt-mod-states} A and B
show that the hole is largely localized on the odd-sites and the electron is localized on the 
even-sites.  The relaxed state is a self-trapped state with the electron largely localized 
on site 6 and the hole distributed on either side on sites 5 and 7.   States C and D are the 
vertical and relaxed odd-parity state corresponding to a dissociated charge-transfer 
state.  Finally, E and F in this figure correspond to the state responsible for the absorption 
peak at 4.4eV.  In this state, the electron and hole are more or less uniformly delocalized over the 
entire polymer segment and carries an oscillator strength comparable to the 2.14eV state, 
again consistent with the semiempirical data.~\cite{jespersen:12613}

\begin{figure}[t]
\includegraphics[width=\columnwidth]{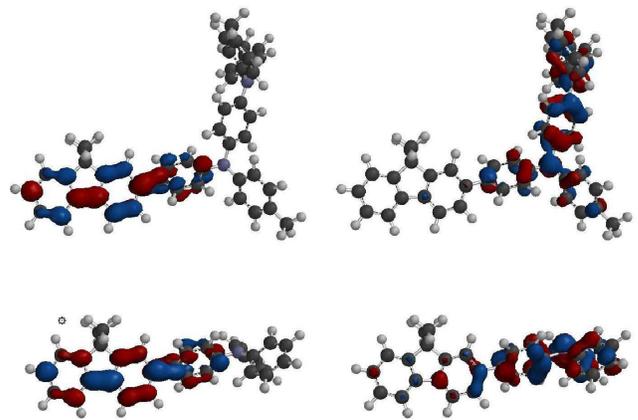}
\caption{(color onlne) Semiempirical (PM3) LUMO (left) and HOMO (right) wave functions for PFB (top) and TFB (bottom) 
monomer}\label{pfb-tfb-orbitals}
\end{figure}

\begin{figure}[t]
\includegraphics[width=\columnwidth]{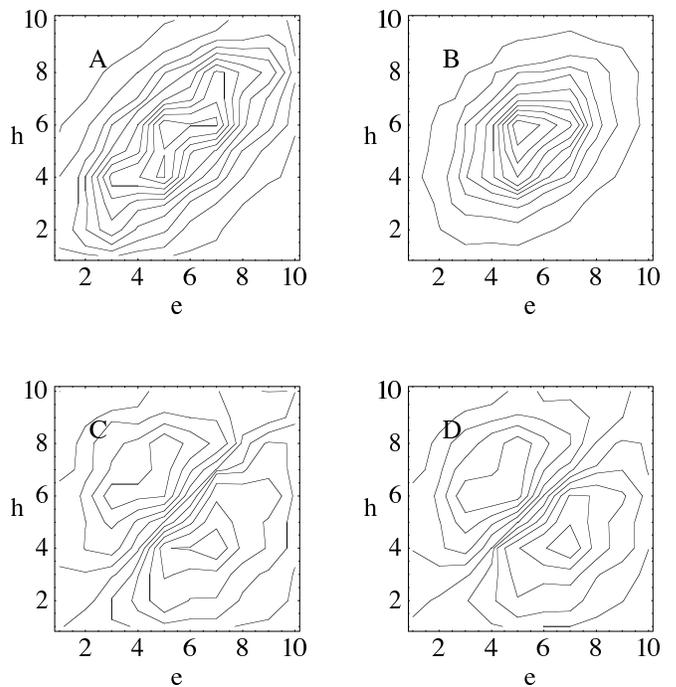}
\caption{Vertical and adiabatic exciton (A,B)  and charge-separated states (C,D) for the 10 site TFB chain. See text for details.}\label{tfb-states}
\end{figure}

For PFB and TFB monomers,  semiempirical (PM3) calculations at the ground state equilibrium configuration 
give HOMO energies of  -8.00748eV and -7.95994 eV and LUMO energies of -0.01862 eV and 0.05201 eV respectively 
for isolated phenylenediamine and diphenylamine units.  In comparison, the semiempirical HOMO and LUMO levels for the
fluorene segment  are -8.83304eV and -0.32915eV, respectively.   In the combined co-polymers, the LUMO 
is more or less localized on the fluorene as in the F8BT case as shown in Fig.~\ref{pfb-tfb-orbitals}.   To account for this 
modulation within our model, we include an 0.2eV modulation
about the band centers listed in Table~\ref{table1} in the 
conduction and valence site energies for the PFB and TFB chains.  
Since the net modulation is on the order of the exciton binding energy, the 
the electron and hole will be localized on alternate repeat units in the lowest excited state as indicated
in Fig.~\ref{tfb-states}-A for the vertical exciton and Fig.~\ref{tfb-states}-B for the 
adiabatic  or self-trapped exciton.   Even though the electron and hole are 
on different lattice sites, they remain a bound pair.  
Fig.~\ref{tfb-states}-C and -D show the dissociated electron/hole pair as evidenced by the 
nodal line along the $e=h$ diagonal.  

Table ~\ref{table3} summarizes the relevant states for the isolated chains discussed in this paper.  
It is interesting to note that for the unmodulated chains (PPV and BBL) the adiabatic 
exciton binding energy is 0.1eV higher than in the modulated co-polymer
 chains (F8BT, TFB,  and PFB).     Having established models for the electronic states
 of the isolated chains, we turn our attention towards modeling the electronic states of the 
 polymer heterojunctions.

\subsection{Electronic states of heterojunctions}
We consider three model heterojunction systems:  PPV:BBL, TFB:F8BT, and PFB:F8BT.  
As discussed in the introduction, devices fabricated using these polymers are either
layered nanostructures or phase-separated polymer blends.  
The crucial  energetic consideration is the exciton destabalization threshold.  
In an ideal non-interaction electron model, 
a type II heterojunction will destabilize an exciton present at the interface 
if the exciton binding energy is less than the band-off set.  In organic semiconductors, 
the Coulombic interaction between the electron and the hole is quite significant with 
exciton binding energies in excess of 0.5 eV in  most materials.  When this 
is on the order of the band off-set, excitons   can be stable at the interface giving 
LED behavior.    By selecting materials with different HOMO and LUMO energies, one can 
effectively tune LED or photovoltaic performance of a heterojunction material. 

\subsubsection{PPV:BBL}

\begin{figure}[t]
\includegraphics[width=\columnwidth]{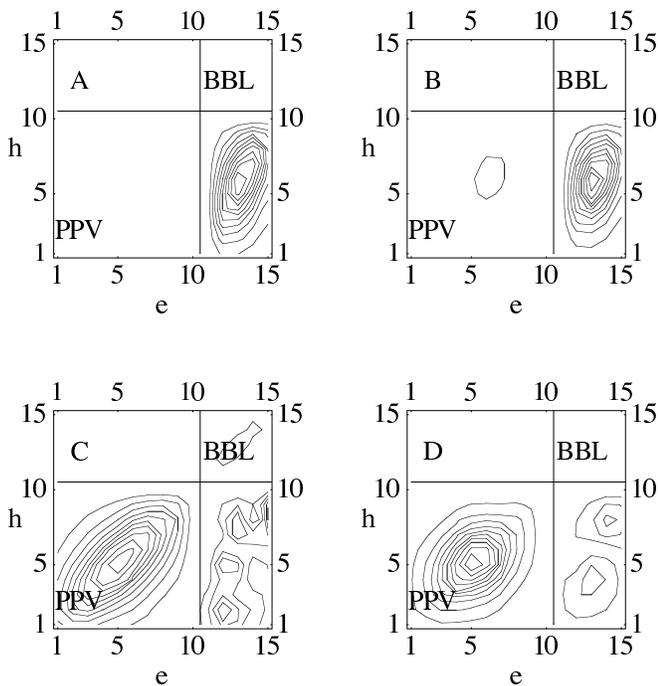}
\caption{
Excited state electron/hole densities for PPV:BBL heterojunction. The axes "e" and "h" refer
to the electron and hole ``coordinates'' on the lattice.   Sites 1-10 are the PPV sites and 
sites 11-15 are the BBL sites.   Plots A and B show the density for the lowest energy
vertical and adiabatically relaxed charge-transfer state while C and D correspond to 
the vertical and relaxed density for the excitonic state indicated in the correlation diagram. 
In C and D one can clearly see the mixing between the intrachain exciton on PPV and
interchain charge-separated configuration with the hole residing on PPV and the
electron being transferred to the BBL.}\label{ppv-bbl-states}
\end{figure}

The first of these semiconductor heterojunction materials we consider here, 
PPV:BBL, has been used in the 
fabrication of solar-cells and photovoltaic devices and a recent 
review by Alam and Jenehke provide a succinct overview of recent progress towards the
fabrication of efficient solar-cell devices using layers of organic polymer donor/acceptor
heterojunction materials.\cite{alam:4647,jenekhe:765,granstrom:257,bach:583,kamohara:014501,russell:2204,jenekhe:2635,chen:487,lin:3495,weng:6838,xue:3013,osaheni:3112,gao:2778,gregg:31,saito:116,yu:5243,jenekhe:2635,manoj:4088,narayan:3938,nakamura:6878, schilinsky:2816}   By alternating nanoscale bilayers of 
high electron affinity (EA) acceptor polymers such as BBL with an EA = 4.0eV with a donor such as PPV or its soluble derivative MEH-PPV
exhibit rapid and efficient photoinduced charge transfer and large photovoltaic power conversions
under white light illumination at 80-100 m/cm$^2$ (AM1.5).\cite{alam:4647}     
Since the
band off-set at the heterojunction of the materials is larger than the exciton binding energy  
in PPV:BBL,  photoexcitation of either BBL or PPV will result in the production of 
a charge-separated interchain species. 

\begin{figure}[t]
\includegraphics[width=\columnwidth]{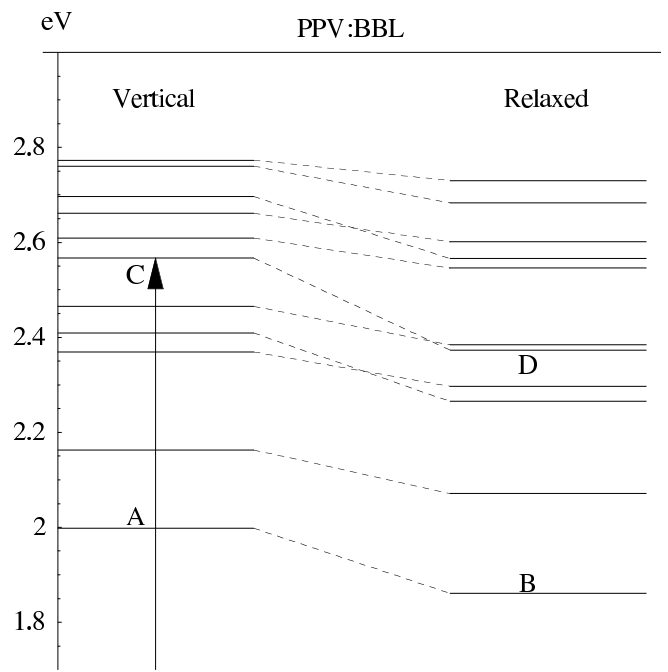}
\caption{Correlation between vertical and adiabatic energy levels for the PPV:BBL heterojunction. 
The vertical scale is the excitation energy.  Labeled states correspond to densities plotted in 
Fig.~\ref{ppv-bbl-states}.}\label{ppv-bbl-correlation}
\end{figure}

The relevant states for the PPV:BBL heterojunction are shown in Fig.~\ref{ppv-bbl-states}.
The top two figures are electron/hole densities for inter-chain charge-separated
state while the bottom two are the vertical and adiabatic densities for the exciton.  
Fig.~\ref{ppv-bbl-correlation} indicates the correlation between the vertical and relaxed 
energy levels.    The vertical exciton (Fig.~\ref{ppv-bbl-states}-C) corresponds to the state
with the largest transition dipole to the $S_o$ ground state.

First, we note that the lowest energy excited state in this system is the charge-transfer state 
in which the hole resides on the PPV chain (sites 1-10 in Fig.~\ref{ppv-bbl-states}) and the 
electron resides on the BBL chain (sites 11-15).  There is some mixing between the interchain 
charge-transfer state and geminate electron/hole configurations on either chain as evidenced by the
smallest contour rings in Fig.~\ref{ppv-bbl-states} A and B.  Clearly, $>$99.9\% of the population is 
in interchain charge-transfer configurations.   The vertical excitonic state, Fig.~\ref{ppv-bbl-states} C, 
is largely localized on the PPV chain with some mixing between the chains.    We find that the
extent of the interchain mixing is determined largely by the interchain hopping integral,  which we set 
at $t_\perp = 0.15 t_{||} = 0.0804eV$.     In the self-trapped exciton (Fig.~\ref{ppv-bbl-states} D), the 
interchain mixing is weaker.

The energy level diagram shown in Fig.\ref{ppv-bbl-correlation} shows the correlation between the 
vertical excited states (at the ground state equilibrium geometry ($Q_\mu = 0$) with the adiabatically
relaxed excited states, each of which has a unique excited state equilibrium geometry.  
This correlation diagram indicates a number of intersystem crossings can occur as the system
relaxes from the vertical to adiabatic geometries.  In particular, notice that 
the excitonic state (C) intersects another state as it relaxes to D.  This state is nearly 
degenerate with the adiabatic exciton (D) and is predominantly an interchain charge-transfer state. 
It is interesting to speculate the role that such intersections and close degeneracies 
will play in the photophysics of this system.   

\subsubsection{PFB:F8BT vs. TFB:F8BT}

\begin{figure}[t]
\includegraphics[width=\columnwidth]{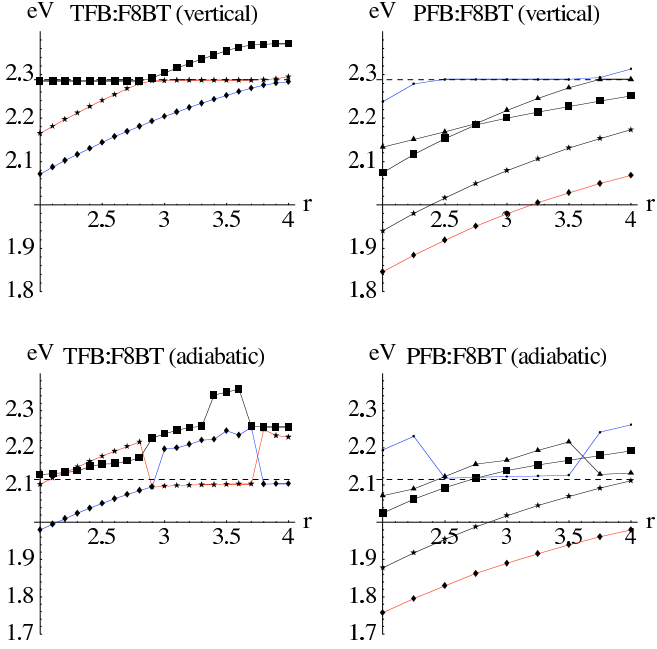}
\caption{Variation of vertical CI energy levels for cofacial PFB:F8BT and TFB:F8BT chains with 
chain separation, $r$, taken units of the lattice constant, $a$.  The horizontal dashed line corresponds to
the exciton energy level for the isolated F8BT chain.  }\label{dist-corel}
\end{figure}

\begin{figure}[t]
\includegraphics[width=\columnwidth]{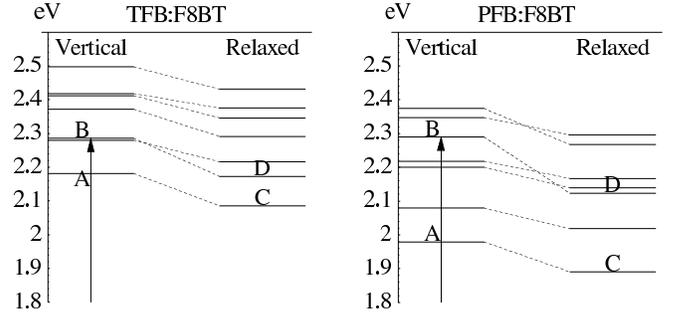}
\caption{Correlation between vertical and adiabatic energy levels for TFB:F8BT and PFB:F8BT heterojunctions.  The vertical scale is the excitation energy in eV. Up and down arrows correspond to states present in absorption or emission spectra of each heterojunction.  
In PFB:F8BT, the lowest energy state is dark.  The levels labeled A-D correspond to the states plotted in Fig.~\ref{tfb-f8bt-states}
and Fig.~\ref{pfb-f8bt-states}. 
 }\label{tfb-pfb-f8bt-correlation}
\end{figure}

As discussed earlier, TFB:F8BT and PFB:F8BT sit on either side of the exciton destabalization 
threshold.    In TFB:F8BT, the band off-set is less than the exciton binding energy and 
these materials exhibit excellent LED performance.  On the other hand, devices
fabricated from PFB:F8BT  where the exciton binding energy is less than the 
off-set, are very poor LEDs but hold considerable promise for photovoltic devices.  In both of
these systems, the lowest energy state is assumed to be an interchain exciplex
as evidenced by a red-shifted emission about 50-80ns after the initial photoexcitation.
\cite{morteani:1708}  In the case of TFB:F8BT, the shift is reported to be 140$\pm$20 meV
and in PFB:F8BT the shift is 360$\pm$30 meV relative to the exciton emission, which originates
from the F8BT phase.   Bearing this in mind, we systematically varied the separation 
distance between the cofacial chains from $r = 2a - 5a$  (where $a$ = unit lattice constant) 
and set $t_\perp = 0.01$
in order to tune the Coulomb and exchange coupling between the chains and calibrate
our parameterization. 
Fig.~\ref{dist-corel} shows the variation of the lowest few diabatic states with $r$ for both 
heterojunctions.   The dashed line in each corresponds to the exciton energy for an 
isolated F8BT chain.  
 For large interchain separations, the exciton remains localized on the F8BT chain in 
 both cases.  As the chains come into contact, dipole-dipole and direct Coulomb couplings 
 become significant and we begin to see the effect of exciton destabilization.   For TFB:F8BT, 
 we select and interchain separation of $r  = 2.8a$ giving a 104meV  splitting between the 
 vertical exciton and the vertical  exciplex  and 87.4 meV for the adiabatic states.  
 For PFB:F8BT,
 we chose $r = 3a$ giving a vertical exciton-exciplex gap of 310 meV and an adiabatic 
 gap of  233 meV. In both TFB:F8BT and PFB:F8BT, the 
 separation produce interchain exciplex states as the lowest excitations. 
with energies reasonably close to the experimental 
shifts. 

\begin{figure}[t]
\includegraphics[width=\columnwidth]{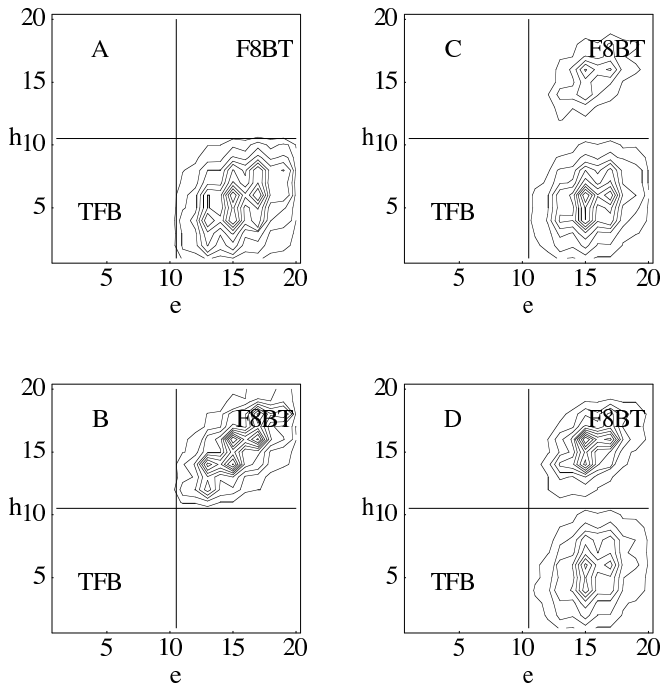}
\caption{
Excited state electron/hole densities for TFB:F8BT heterojunction.  
The electron/hole coordinate axes are the same as in Fig.~\ref{ppv-bbl-states}
except that sites 1-10 correspond to TFB sites and 11-20 correspond to F8BT sites.
 Note the weak mixing between the 
interchain charge-separated states and the F8BT exciton in each of these plots.
}\label{tfb-f8bt-states}
\end{figure}

\begin{figure}[t]
\includegraphics[width=\columnwidth]{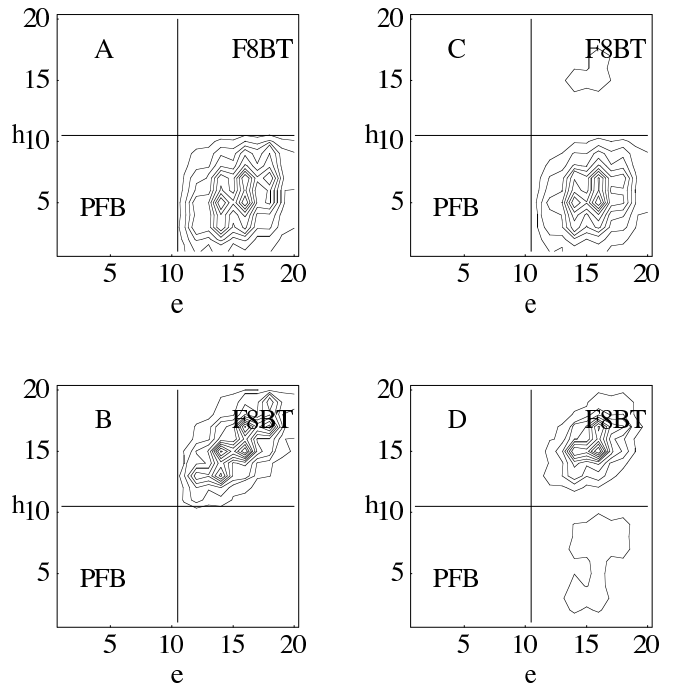}
\caption{
Excited state electron/hole densities for PFB:F8BT heterojunction.
The axes are as in previous figures  except that sites 1-10 correspond to PFB
sites and sites 11-20 to F8BT sites. }\label{pfb-f8bt-states}
\end{figure}

Fig.~\ref{tfb-pfb-f8bt-correlation}  compares the vertical and adiabatic energy levels in the TFB:F8BT and PFB:F8BT
chains and Figs.~\ref{tfb-f8bt-states} and \ref{pfb-f8bt-states} show the vertical and relaxed exciton and charge-separated states
for the two systems.  Here, sites 1-10 correspond to the TFB or PFB chains and 11-20 correspond to the
F8BT chain.
The energy levels labeled in Fig.~\ref{tfb-pfb-f8bt-correlation}
correspond to the states plotted in Figs.~\ref{tfb-f8bt-states} and \ref{pfb-f8bt-states}.  
We shall refer to states A and B as the vertical exciplex  and vertical  exciton and to states
C and D as the adiabatic exciplex and adiabatic exciton respectively. 
Roughly, speaking a pure exciplex state
will have the charges completely separated between the chains and will contain 
no geminate electron/hole configurations.  Likewise, strictly speaking, 
a pure excitonic state will be localized to 
a single chain  and have only geminate electron/hole configurations.  
Since site energies for the the  F8BT chain are modulated to reflect to 
internal charge-separation in the F8BT co-polymer as discussed  above, 
we take our ``exciton''  to be the lowest energy
 state that is localized predominantly along the diagonal in the F8BT ``quadrant''.

In the TFB:F8BT junction, the lowest excited state is the exciplex for both the vertical and 
adiabatic lattice configurations with the hole on the TFB and the electron on the F8BT. 
(Fig.\ref{tfb-f8bt-states}A,C)   In the vertical case,  there appears to be very little coupling between 
intrachain and interchain configurations.  However,  in the adiabatic cases there is 
considerable mixing between intra- and inter-chain configurations.   First, this 
gives the adiabatic exciplex an increased transition dipole moment to the ground state. 
Secondly,  the fact that the adiabatic exciton and exciplex states are only 87 meV apart means that
at 300K, about 4\% of the total excited state population will be in the adiabatic exciton.

For the PFB:F8BT heterojunction, the band off-set is greater than the exciton binding energy and sits squarely 
on the other side of the stabilization threshold.   Here the lowest energy excited state
(Figs.~\ref{pfb-f8bt-states}A and B) 
 is the interchain charge-separated
state with the electron residing on the F8BT (sites 11-20 in the density plots in Fig~\ref{pfb-f8bt-states})
and the hole on the PFB (sites 1-10).  The lowest energy exciton is almost identical to the exciton in the 
TFB:F8BT case.   Remarkably, the relaxed exciton (Fig.~\ref{pfb-f8bt-states}D) shows slightly more interchain
charge-transfer character than the vertical exciton (Fig.~\ref{pfb-f8bt-states}C).  
While the system readily absorbs at 2.3eV creating a localized exciton on the F8BT,  luminescence 
is entirely quenched since all population within the excited states is readily transfer to the lower-lying
interchain charge-separated states with vanishing transition moments to the ground state.

\begin{figure}[b]
\includegraphics[width=\columnwidth]{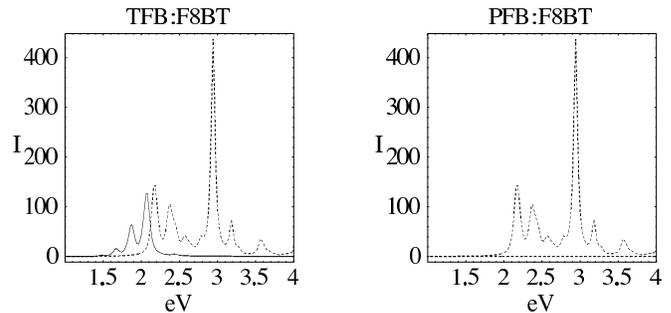}
\caption{ 
Vibronic absorption (dashed) and emission (solid)  spectra of TFB:F8BT and PFB:F8BT
heterojunctions.  Note that in the TFB:F8BT the fluorescence spectrum is scaled by a 
factor of 100.}\label{tfb-pfb-spectra}
\end{figure}

The mixing between the exciplex and exciton in TFB:F8BT can be seen in the 
predicted emission spectra for the system. Fig.~\ref{tfb-pfb-spectra}  shows the
absorption and emission spectra for the TFB:F8BT and PFB:F8BT computed by taking the
transition dipole moment between the vertical  and relaxed excited states to the 
ground state.  While the fluorescent emission from TFB:F8BT is weaker than the
corresponding absorption, it is not entirely quenched as in the PFB:F8BT case.

\section{Discussion}

In this paper,  we developed a model for donor-acceptor semiconducting polymer  heterojunctions.  
For the PPV:BBL heterojunction, the band off-sets at the polymer interface is sufficient to 
dissociate an electron/hole pair created by photoexcitation.  Because of this natural internal 
driving force to create charge-separated states, PPV:BBL blends are of considerable interest to 
materials chemists and device engineers developing efficient photovoltaic cells for solar energy conversion~\cite{alam:4647}.
TFB:F8BT and PFB:F8BT are very similar heterojunctions.  Within our model, the single 
parametric difference between 
TFB and PFB is the location of the band-center for the valence band.  In TFB:F8BT, the off-set between the 
valence bands is (-2.98eV+ 3.54eV)  0.56eV which is right at the exciton binding energy.  Hence, we see considerable 
mixing between the intrachain exciton localized on the F8BT and a charge-separated state.  
In the PFB-F8BT case, the valence band off-set is 0.79 eV, clearly greater than the exciton binding energy. 
Hence, the lowest energy excited states are predominantly interchain charge-separated states. 

In a forthcoming paper, \cite{bittner05b}, we continue along the lines set forth in this paper by computing the 
state-to-state relaxation dynamics following both photoexcitation and electron/hole injection. 
We will compare results obtained within a vertical approximation (whereby transitions occur
via phonon creation/annihilation, but the lattice is frozen in the ground-state equilibrium geometry)
to those obtained within an adiabatic Marcus-Jortner-Hush approximation where the transitions
occur between the equilibrium geometry of each excited state.  
We will also discuss the role that the avoided crossings and conical intersections of the 
Born-Oppenheimer potential energy surfaces of the excited states play in the electron/hole 
capture process, exciplex formation, and in the thermal regeneration of intramolecular 
excitons in TFB:F8BT.



\begin{acknowledgments}
This work was supported in part by the National Science Foundation and the Robert A. Welch Foundation.  
\end{acknowledgments}


%




\bibliography{Photovoltaics}

\end{document}